\documentclass{article}
\usepackage[final]{neurips_2020}
\usepackage[utf8]{inputenc}
\usepackage[T1]{fontenc}
\usepackage{hyperref}       % hyperlinks
\usepackage{url}            % simple URL typesetting
\usepackage{booktabs}       % professional-quality tables
\usepackage{amsfonts}       % blackboard math symbols
\usepackage{nicefrac}       % compact symbols for 1/2, etc.
\usepackage{microtype}      % microtypography
\usepackage{natbib}
\usepackage{graphicx}
\defcitealias{HowardResearch20:2020}{Targeted Infusion: Engaging Howard University Computer Science Students in Interactive Computing Infused Curricula}
\defcitealias{HowardAffective}{Affective Biometrics Lab}

\title{Towards an Abolitionist AI: the role of Historically Black Colleges and Universities}
\author{%
  Charles C.~Earl \thanks{All opinions and analysis in this work are the author?s alone.}\\
  Automattic.com\\
  \texttt{charles.earl@automattic.com}
}

\begin{document}

\maketitle

\begin{abstract}
\emph{Abolition} is the process of destroying and then rebuilding the
structures that impede liberation. This paper addresses
the particular case of Black folk in the United
States, but is relevant to the global decolonization movement. Using notions of 
abolition and infrastructures of feeling developed by  Ruth Wilson Gilmore, I view \textbf{Historically Black Colleges and Universities} ( \emph{HBCUs} ) as a particular kind of abolitionist project, created for the explicit purpose of nurturing and sustaining  Black excellence particularly within the sciences. I then examine how artificial
intelligence (AI) in particular and computing in general have contributed
to racial oppression and the further confinement and diminishing of Black existence. I conclude by examining how the space held by HBCUs in computing
might contribute to a reimagining of AI  as a technology that enhances the possibility and actualization of Black life. 
\end{abstract}

\section{Introduction}

As a child, I attended the nursery school of Spelman College, a historically Black women's college located in 
Atlanta. One of my earliest memories is of witnessing a protest taking place 
on the campus. The young women of Spelman, in their proud defiant afros, were protesting the 
visit of Nelson Rockefeller. They were demanding that the Rockefeller family -- a major funder of the college -- divest 
from all assets invested in the South African apartheid regime. Rockefeller's grandfather, John D Rockefeller, had
given the college its first major endowment, and what had been the
Atlanta Baptist Female Seminary had been renamed in honor of John Rockefeller's wife, Laura
Spelman. 

In an age that demands resistance and the abolition of structures that
continue to oppress Black people, what is the role of historically Black institutions
like Spelman College? And in particular what should be the role of those institutions in defining how Black people 
interact with and define the development of AI? Given the outsize role these institutions play in the development of 
Black computing professionals -- 9\% of Black undergrads attend HBCUs, yet  
HBCUs produce 37\% of the Black STEM graduates in the United States -- it is important to assess how these graduates are  prepared to address the inequities in computing that exacerbate these harms. Further, these institutions continue to receive significant funding from 
sources whose aims are at odds with liberation. In this paper,  my hope is to raise questions and initiate dialog about
how these institutions might contribute to liberatory computing.

\section{Abolition and the geographies of freedom}
\label{sec:orgbbc76fc}

%\begin{quote}
%\textit{Capitalism requires inequality and racism enshrines it} 
%\\
%-- Ruth Wilson Gilmore, \citeyearpar{ruthieInnocence}
%\end{quote}

The geographer Ruth Wilson
Gilmore has been one of the most eloquent and incisive voices
articulating what abolition means in 2020. Her book \textit{The Golden Gulag} \cite{golden_gulag}  is both a meditation on the
campaign to slow the growth of state and private prisons in California
as well as an exploration of the harm done by removing scores of people from
society for the sake of retribution. In the specific case of prisons,
Gilmore is asking us to confront what we think prisons are for -- punishment? rehabilitation? justice? -- and the
reality of what they actually accomplish.

Though Gilmore's work focuses primarily on prison
abolition, she articulates a broad, encompassing vision of liberation, developed out of her geographer's view of the world. In her essay \textit{Abolition Geography and the
Problem of Innocence} \cite{ruthieInnocence} she writes \textit{Abolition geography starts from the homely premise that \textbf{freedom is a place} } \citep[pg.~227]{ruthieInnocence}.  Abolition then is overall the affirming of life out of places intended
for domination, exploitation, and ultimately spiritual, intellectual
and physical death:

\begin{quote}
If unfinished liberation is the still-to-be-achieved work of
abolition, then at the bottom what is to be abolished isn't the past
or its present ghost, but rather the process of hierarchy,
dispossession, and exclusion that congeal in and as
group-differentiated vulnerability to premature death.  \citep[pg.~228]{ruthieInnocence}
\end{quote}

It is then this process of dispossession of life itself -- epitomized by prisons -- that is the
enforcement of the inequality necessitated by capitalism. The central tenet of Gilmore's work is the examination of these
qualities of existence. The carceral landscape of capitalism that
demands an extractive hierarchy on the one hand, and the affirming alternate reality that the
exploited imagine and create in real time and in real spaces on the other. 

But Wilson's work sees liberation as a dual of abolition. Referencing W.E.B. DuBois's  classic \textit{Black
Reconstruction} \citeyear{DuBois} she makes clear that the formerly enslaved were always creating free spaces in which they 
could exist as whole human beings. For Wilson, this continuous building creates an infrastructure of feeling \citep[pg.~236-7]{ruthieInnocence} -- an emotional, spiritual practices that connect the present generation to the love and aspirations of ancestors that struggled and dream of independence. 

HBCUs exemplify a place of freedom in Gilmore's conception -- most having been birthed in during Reconstruction. HBCUs provide havens in which Black scholars can exist without fear of thinking/coding while Black. In his study of the affirming environment at Florida A\&M University (one of the largest producers of Black CS grads) Allen concludes that HBCUs
"provide a relief from the burdens of oppressive articulations and experiences of society and space" \cite{black_geographies}. 

The impact that this space cannot be overstated. There are 102 HBCUs in the U.S. Despite the fact that 9\% of Black undergrads students attend these institutions, they account for 37 percent of the degrees awarded to Black people in STEM disciplines. 33 percent of the Black recipients of doctoral degrees in computing earned their first degree from a HBCU,  and 12 \% of the Black PhDs in computing receive their doctorate from an HBCU. Over the nine year period 2007 - 2017, the HBCUs Spelman College, Howard University, and North Carolina A\&T State University had the highest yield of doctorates (Black CS or Math doctorates per 100 Bachelors degrees in STEM) of any institutions in the United States \cite{HBCUImpact}. Beyond the CS discipline, two of the most impactful thinkers on abolition -- Dr. Ruha Benjamin and Dr. Tressie McMillan Cottam -- themselves received undergraduate degrees from HBCUs -- respectively Spelman College and North Carolina Central University.

\section{The need for abolition in AI}
\label{sec:org002149c}

If capitalism requires inequity, and if racism enshrines it  \cite{ruthieInnocence}, then the
hyper profit motive of AI applied to platform capitalism has been a successful effort to optimize racism. The uses of AI -- particularly in platforms that derive value from manipulative practices -- are necessarily not in the interest of Black people. \cite{ruha_jim_code}, \cite{gender_shades} and others have
documented how AI has been used to enforce racial hierarchies of power. \cite{Cottom2020WherePC} further discuses how the 
logic of racial capitalism has been ensconced into the logic of all web scale platforms -- it is at this point core to Internet economy.

From above work we can distill a few kinds of technologies and practices that reinforce racial oppression:
  
\begin{itemize}
\item  \emph{Carceral technologies}: for example predictive policing and algorithmic sentencing are used explicitly for the purposes of capturing, isolating, and destroying Black life.
\item  \emph{Discriminatory resource allocation} existing technologies are used to limit the access of Black people to resources
from health care to credit, often optimizing for profit over well-being.
\item  \emph{Surveillance technologies} are used explicitly for expropriation of behavioral surplus and circumscribing autonomy by predictively tracking and controlling the behavior of the surveilled.
\item  \emph{Racialized hiring and software engineering practices} practices of recruiting, advancement, and project management within software organizations that enable the reproduction of racial harms.
\item \emph{Militarization of AI} particularly in the secretive use of technology to further colonialist agendas on the global south -- the use of this technology upon the civilian Black and Indigenous population of the U.S.  is particularly concerning. 
\end{itemize}

On the other hand, there are tools that Benjamin identifies as being able to dismantle these technologies, mitigate their harm, or even better create the spaces that allow free spaces to come into being.

\begin{itemize}
\item \emph{Creating space for  and voice to the marginalized}, the practice of centering marginalized voices by enabling full participation, contribution, and equitable control of the development process.
\item  \emph{Algorithmic and process audits} that identify the harms caused by algorithmic decision making and also identify 
the systemic issues in the software development process that allow them to go unchecked.
\item \emph{Racial abuse protection and mitigation} technologies that intercept racial attacks and mitigate harm (e.g. micro-aggression classifiers)
\item  \emph{Liberatory design} practices which enable the development of software products center the aspirations and 
requirements of marginalized community members.
\item  \emph{Equitable work practices} that recognize and reward the contribution of data annotators, content moderators, and other "gig" workers whose value is extracted and erased by racialized platform capitalism.
\end{itemize}

\section{Abolition and the reproduction of oppression at HBCUs}
\label{sec:orgca89d72}

We can get some understanding of the degree to which abolitionist computing within HBCUs occurs by looking at the 
research funding. The major U.S. federal agencies -- among them Department of Defense (DoD), National Institutes of Health (NIH), and National Science Foundation (NSF) -- allocate large grants specifically targeted to improve the research and teaching capacity of HBCUs. The NSF annually funds the HBCU Excellence in Research (EiR) program (\$10 million per year), and the Historically Black Colleges and Universities Undergraduate Program (HBCU-UP) (\$55 million per year). 

A close look at research activity at Howard University raises the other question of to what degree Black computing scholars are being trained to participate in the reproduction of racist technology. At Howard, approximately \$15 million awarded in computing research for 2020 (excluding on-going grants), 50\% were from department of defense grants in bio-metrics and cyber defense (e.g. Howard was awarded \$7.5 million for a DoD Center for Excellence in AI/ML \cite{DoDEx:2020} ), 4\% from educational awards for development of retention and STEM curricula, while 1\% was awarded to research specifically aimed at addressing 
online abuse directed toward Black people (e.g. "Deep Learning of Sentiment Analysis and Codeswitching for Identification of Harmful Technologies" and other initiatives from Dr. Gloria Washington's  \citetalias{HowardAffective} ). Similar patterns emerge when looking at other HBCUs like North Carolina A \& T State University.

That is, with heavy DoD and platform capitalism support Black 
students and researchers must navigate being participants in the reproduction of what \cite{DecolonialAI} refers to as colonizing AI especially with respect
to surveillance and military technology. This is concerning especially given the potential harms for Black folk in the U.S. and global south \cite{militaryAI}.

\section{Abolitionists agendas for HBCUs}
\label{sec:org440017f}

To begin in earnest the development of abolitionist approaches to AI within the HBCU landscape requires a return to the 
source. Tapping into the infrastructure of feelings that already permeates these institutions and enables their success in nurturing 
so many impactful people. Within available resources already devoted to capacity development, there could be informal meetups, conferences, gatherings across schools, across disciplines to explore the implications of abolitionist approaches to computing. The significant number of students involved in BLM actions combined with resources already devoted to making computing accessible could be a powerful accelerator.  The development of conferences and centers to promote tools for community audits and interrogation of software platforms is another approach that could have wide adoption and immediate impact given the  millions of Black people are impacted daily by racialized ad targeting and unfair pricing experiments by software platforms. Lastly, the Association for Computing Machinery (ACM)  has long acknowledged the dearth of adequate ethics education in computing. For AI in particular, this is especially necessary given the potential for harm. Perspectives on ethics that would explore methodologies for assessing what should and should not be built would be an important preparation for students. Engagement with scholars of the social sciences, law, and healthcare -- present at so many HBCUs -- that could collectively work on the development of archives and models capable of addressing community needs is also important.

\section{Conclusion}
\label{sec:org7a2552a}

I have attempted to consider the current state of affairs in AI viewed from the perspective
 of abolition as articulated by Ruth Wilson Gilmore. Viewing AI as a carceral space that replicates and accelerates 
 inequity I have attempted to look at the way in which  HBCUs might chart a new course for it. There remains much to be done.

\bibliography{abolish}

\end{document}